# Masses of $^{130}$Te, $^{130}$Xe and double beta-decay $Q$-value of $^{130}$Te

Matthew Redshaw[1], Brianna J. Mount[1], Edmund G. Myers[1], and Frank T. Avignone III[2]

[1]*Department of Physics, Florida State University, Tallahassee, Florida 32306-4350, USA*

[2]*Department of Physics and Astronomy, University of South Carolina, Columbia, South Carolina 29208, USA*

**The atomic masses of $^{130}$Te and $^{130}$Xe have been obtained by measuring cyclotron frequency ratios of pairs of triply-charged ions simultaneously trapped in a Penning trap. The results, with one standard deviation uncertainty, are $M(^{130}\text{Te}) = 129.906\ 222\ 744(16)$ u and $M(^{130}\text{Xe}) = 129.903\ 509\ 351(15)$ u. Allowing for cancellation of systematic errors in the mass difference, the double-beta-decay $Q$-value, required for searches for the neutrino-less double-beta-decay of $^{130}$Te, is determined to be $Q_{\beta\beta}(^{130}\text{Te}) = 2527.518(13)$ keV.**

There are important questions in neutrino physics that are most effectively addressed by neutrino-less double-beta ($0\nu\beta\beta$) decay experiments [1-3]. Are neutrinos Majorana particles, *i.e.* particles that differ from antineutrinos only by helicity? If yes, then $0\nu\beta\beta$−decay can occur, and violate lepton-number conservation. What is the neutrino mass-scale? Measurement of the oscillations of atmospheric neutrinos [4] and the confirmation of the oscillations of the chemical solar neutrino experiments [5-7] by SuperKamiokande [8], and the results of the SNO experiment [9] that showed that the predicted flux of $^8$B solar neutrinos [10] was correct, together with the KamLAND reactor-neutrino experiment [11] – which gave clear evidence that the Mikheyev-Smirnov-Wolfenstein (MSW) large mixing-angle solution of solar neutrino oscillations is the strongly favored one [10] – imply scenarios in which the "effective Majorana mass of the electron neutrino", $m_{\beta\beta}$, could be larger than 0.05 eV/$c^2$. Detectors that are planned or under construction could allow the observation of $0\nu\beta\beta$−decay at this scale.

To date, the most sensitive limits on $m_{\beta\beta}$ have come from experiments searching for the $0\nu\beta\beta$−decay of $^{76}$Ge using germanium semi-conductor detectors [12,13], and large-scale $^{76}$Ge experiments [14,15] are under development. One sub-group has even claimed an observation [16,17], although this has been disputed [18]. Nevertheless, due to both uncertainties in the nuclear matrix elements as well as difficulties in confirming the observations, searches for $0\nu\beta\beta$−decay in more than one nucleus are required. A competitive, high-resolution experiment seeks to detect $0\nu\beta\beta$−decay in $^{130}$Te, by using cryogenic bolometers consisting of single crystals of TeO$_2$ bonded to high-sensitivity germanium thermistors. A prototype version, CUORICINO [19], was operated until July 2008 at the Laboratori Nazionali del Gran Sasso (LNGS), while a full-scale version, CUORE [20], which will contain approximately 750 kg of TeO$_2$ or ~200 kg of $^{130}$Te, is under construction.

A crucial datum for $0\nu\beta\beta$−decay searches is the $Q$-value, the mass-energy difference between the parent and daughter atoms. This defines the location of the expected sharp peak in the sum-energy-spectrum of the two electrons emitted, which is the signature of the neutrino-less decay. In the case of the $^{76}$Ge experiments, the mass-difference measured by Douysset et al. [21] is essential to the interpretation. A mass measurement of $^{136}$Xe by Redshaw et al. [22] will play a similar role in large-scale $^{136}$Xe $0\nu\beta\beta$−decay experiments under development [23]. In the case of $^{130}$Te, the global atomic mass evaluation (AME) gives $Q_{\beta\beta}(^{130}\text{Te}) = 2530.3(2.0)$ keV [24], while an earlier determination by the Manitoba group gave 2528.8(1.3) keV [25]. This uncertainty in the *Q*-value already impacts the analysis of CUORICINO, and would be a serious limitation for the future CUORE, which has an anticipated FWHM energy resolution of 5 keV and absolute energy uncertainty of better than 0.4 keV. Here we report high-precision, cryogenic-Penning-trap measurements of ratios of cyclotron frequencies, yielding the mass-energy difference $[M(^{130}\text{Te}) - M(^{130}\text{Xe})]c^2 = Q_{\beta\beta}(^{130}\text{Te})$ with a one-standard-deviation uncertainty of 13 eV, more than sufficiently precise for all proposed $^{130}$Te $0\nu\beta\beta-$decay detector developments. We also report absolute atomic masses of $^{130}$Te and $^{130}$Xe. Such precise atomic masses provide the "backbone" of reference masses for global mass evaluations of stable and unstable isotopes such as the AME [24]. This is also the first reported measurement of cyclotron frequency ratios using pairs of single, multi-charged ions, simultaneously trapped in a Penning trap.

*Method:* Most of our techniques and our Penning trap mass spectrometer, which was originally developed at MIT [26-28], have been described elsewhere [22,29-33]. Here we give an overview and indicate the developments required for the present measurements. A comprehensive review of precision mass spectrometry is given in ref. [34]. The Penning trap consists of three hyperboloidal electrodes, the ring and two end-caps, which produce a cylindrically-symmetric quadratic electrostatic potential. (An additional set of "guard-ring" electrodes is used to null the lowest-order field distortion parameter, known as $C_4$.) The electrodes are housed inside an ultra-high vacuum insert, submerged in the LHe-filled bore of a carefully shimmed 8.5 T superconducting magnet. The combination of uniform magnetic field and quadratic electrostatic potential results in three harmonic motions for an ion in the Penning trap: the "trap-modified" cyclotron, axial, and magnetron modes, with frequencies $f_{ct}$, $f_z$, and $f_m$, respectively. In the limit of small mode amplitudes, and with no other forces on the ion, the "true" cyclotron frequency, defined by $f_c = qB/2\pi m$, is given *exactly* by the Brown-Gabrielse invariance theorem, $f_c^2 = f_{ct}^2 + f_z^2 + f_m^2$ [35]. In our Penning trap, only the axial motion is detected directly (and also damped) by interaction with a self-resonant superconducting inductor with a *Q* of 33,000 and center frequency near 213 kHz, coupled to a dc-SQUID.

Single $^{129,130,132}$Xe$^{3+}$ and $^{130}$Te$^{3+}$ ions were made inside the trap by electron-impact ionization of neutral atoms entering through a small hole in the upper end-cap electrode. For the xenon isotopes, small quantities of gas were admitted at the top of the cryogenic insert, approximately 2m above the trap. In the case of tellurium, we injected vapor using an electrically heated dispenser containing a few mg of $^{130}$Te powder at the top of the insert. Unwanted ions were removed by exciting their axial motion, and then lowering the potential of the lower end-cap until the ions struck it. In all cases, making single ions of the desired isotope was facilitated by using samples with more than 90% isotopic enrichment.

The use of higher charge-states increases the signal size for a given axial amplitude, and also increases the cyclotron frequency, both of which improve statistical precision. Smaller mode amplitudes are important in reducing systematic shifts to the mode frequencies due to electrostatic and magnetic field imperfections. These shifts increase with the square or higher powers of the mode amplitudes. Further, the fractional precision with which $f_z$ must be measured, consistent with a given fractional precision for $f_c$ from the invariance theorem, varies as $(f_z/f_c)^2$, favoring high $f_c$. Hence an important innovation with respect to previous work on $^{136}$Xe [22] and $^{129,132}$Xe [29,30] was to extend to these heavy multi-charged ions a "two-ion technique" we had previously developed for singly-charged ions such as $^{28}$Si$^+$ and $^{31}$P$^+$ [31-33]. In this technique, instead of only trapping a single ion at a time, the two ions whose cyclotron frequencies are being compared are simultaneously trapped: one ion is at the center of the trap where its cyclotron frequency is measured, while the other is temporarily "parked" in a large radius cyclotron orbit. The ions are then interchanged, the cyclotron frequency of the new inner ion is measured, and so on. Since the interchange time is typically 5-10 minutes, this enables many more interchanges in a run-time of up to 15 hours (limited by the ion lifetime or the need to refill a LN$_2$ dewar), than the procedures of refs. [22,28-30] in which alternation between the ion species required remaking and isolating each ion, every interchange. This advantage of increased rate of interchange, essential for reducing uncertainty in the cyclotron frequency ratio due to variation in the magnetic field, is obviously greater for ions that are difficult to make. Further, particularly for vapor injection, with possible heating of the Penning trap by thermal radiation and worsening of the trap vacuum by other gases released, it ensures uniform trap conditions for repeated measurements on the two different ions, for as long as they survive against collision with background gas. This can be several days for triply-charged ions (the tellurium data was obtained with a total of three $^{130}$Te$^{3+}$ ions) or several weeks for singly-charged ions.

The actual measurement of the cyclotron frequency of the inner ion used the so-called "pulse-and-phase" (PNP) technique. In brief, after damping all three modes, the cyclotron motion is excited with an electric-dipole rf pulse near $f_{ct}$. Its phase is then allowed to evolve for a variable period of 0.2 to 58 sec, after which the final phase is coherently mapped onto the axial mode, using a quadrupolar "pi-pulse" at the cyclotron-to-axial coupling frequency, $f_{cc} = f_{ct} - f_z$ [26-30]. The evolved cyclotron phase – which by varying the phase evolution time gives $f_{ct}$ – and the axial frequency, $f_z$, are then determined from the ring-down signal of the axial motion following the pi-pulse. The magnetron frequency, $f_m$, is then obtained from $f_z$ and $f_{ct}$ and a measurement, before the run, of the "trap-tilt parameter" [29,30]; $f_c$ is then determined from $f_{ct}, f_z$ and $f_m$ using the invariance theorem. For the measurements used to produce the final cyclotron frequency ratios, the radius of the inner ion's cyclotron orbit was approximately 60 μm, while the parking radius was close to 2 mm.

*Cyclotron frequency ratio measurements:* The $^{130}$Te – $^{130}$Xe mass difference can be obtained from the cyclotron frequency ratio $^{130}$Te$^{3+}$/$^{130}$Xe$^{3+}$. However, instead of measuring this ratio directly, we chose to obtain it from the ratio of the two ratios $^{130}$Xe$^{3+}$/$^{129}$Xe$^{3+}$ and $^{130}$Te$^{3+}$/$^{129}$Xe$^{3+}$. The reason for this is that, although most systematic errors are reduced when comparing ions of similar mass-to-charge ratio, with two ions in the trap, when their mode frequencies are very close (as is the case for the pair $^{130}$Te$^{3+}$/$^{130}$Xe$^{3+}$ with fractional mass difference ~2 x 10$^{-5}$) the ions can no longer be manipulated independently: as we observed, the radial drives which resonantly interact with the inner ion can also excite the outer ion. The resulting systematics, though expected to decrease rapidly with increasing parking radius, are complicated and require further investigation. On the other hand, when measuring $^{130}$Xe$^{3+}$ and $^{130}$Te$^{3+}$ against $^{129}$Xe$^{3+}$ these particular effects are negligible, while other systematics largely cancel in the ratio-of-ratios. Additionally, the comparison with $^{129}$Xe, whose atomic mass (along with that of $^{132}$Xe) we have measured previously to better than 0.1 ppb [30], enables the absolute masses to be determined. Nevertheless, we did perform one run where we directly measured $^{130}$Te$^{3+}$/$^{130}$Xe$^{3+}$ with both ions in the trap; we also took a small amount of data using the simpler procedure in which there is only one ion in the trap at a time – but consisting of only a single set of three $f_c$ measurements on $^{130}$Xe$^{3+}$ followed by a set of three $f_c$ measurements on $^{130}$Te$^{3+}$. To provide further checks of our methods and help estimate uncertainties we also measured the ratio $^{132}$Xe$^{3+}$/$^{130}$Xe$^{3+}$ and the previously measured ratio $^{132}$Xe$^{3+}$/$^{129}$Xe$^{3+}$ [29,30]. Our results for the cyclotron frequency ratios averaged over repeated runs, along with estimated systematic corrections and uncertainties, are given in Table 1. As can be seen, the three different methods for obtaining the $^{130}$Te$^{3+}$/$^{130}$Xe$^{3+}$ ratio gave results consistent within their errors.

TABLE I. Average cyclotron frequency (*i.e.* inverse mass) ratios and systematic corrections for each ion pair. *N* is the number of runs included in the average. $\Delta_{trap}$, $\Delta_{i\text{-}i}$, and $\Delta_{fz}$ are the estimated systematic corrections in parts-per-trillion (ppt), with estimated uncertainty in parentheses, due to trap field imperfections, ion-ion interaction, and shifts in $f_z$ due to ion-detector interaction and differential voltage drift, respectively. $\sigma_{syst}$ is the total systematic error and $\sigma_{stat}$ is the statistical error (in ppt) for each average ratio. <R> is the average ratio after applying systematic corrections, with statistical and systematic uncertainties combined in quadrature, in parentheses. The three entries for $^{130}$Te$^{3+}$/$^{130}$Xe$^{3+}$ correspond to results obtained with a single ion in the trap, with two ions in the trap, and from the ratio of the $^{130}$Te$^{3+}$/$^{129}$Xe$^{3+}$ and $^{130}$Xe$^{3+}$/$^{129}$Xe$^{3+}$ ratios, respectively.

| Ion pair | N | $\Delta_{trap}$ | $\Delta_{i\text{-}i}$ | $\Delta_{fz}$ | $\sigma_{syst}$ | $\sigma_{stat}$ | <R> |
|---|---|---|---|---|---|---|---|
| $^{130}$Xe$^{3+}$/$^{129}$Xe$^{3+}$ | 5 | 1(18) | 1(11) | -18(31) | 38 | 73 | 0.992 311 669 329(82) |
| $^{130}$Te$^{3+}$/$^{129}$Xe$^{3+}$ | 3 | -5(17) | 1(11) | -11(30) | 36 | 75 | 0.992 290 942 332(83) |
| $^{132}$Xe$^{3+}$/$^{130}$Xe$^{3+}$ | 5 | -5(34) | 2(22) | -35(34) | 53 | 83 | 0.984 832 390 737(98) |
| $^{132}$Xe$^{3+}$/$^{129}$Xe$^{3+}$ | 6 | -8(45) | 2(33) | -22(38) | 68 | 65 | 0.977 260 673 493(94) |
| $^{130}$Te$^{3+}$/$^{130}$Xe$^{3+}$ (1 ion) | 1 | -7(26) | 0(0) | 34(15) | 30 | 252 | 0.999 979 112 310(254) |
| $^{130}$Te$^{3+}$/$^{130}$Xe$^{3+}$ (2 ion) | 1 | 2(6) | 0(60) | 0(16) | 62 | 182 | 0.999 979 112 415(192) |
| [$^{130}$Te$^{3+}$/$^{130}$Xe$^{3+}$] ($^{129}$Xe$^{3+}$) | | -6(11) | 0(2) | 7(13) | 17 | 97 | 0.999 979 112 412(98) |

*Systematic corrections and error estimates:* Under the heading $\Delta_{trap}$ in Table I we list estimates of the small corrections that we apply to the observed cyclotron frequency ratios to allow for imperfections in the quadratic electrostatic potential (quantified by $C_4$, $C_6$, etc), and the uniform magnetic field ($B_2$) of the trap, and also due to special relativity [29,30,35]. These field imperfections lead to amplitude-dependent shifts in $f_{ct}$ and $f_z$, the largest being to $f_z$ after the pi-pulse. Differential shifts that affect the ratio occur if either the mode amplitudes, or the values of $C_4$, are different for the two ions. The parameters $B_2$, $C_4$ and $C_6$ were obtained from auxiliary measurements of $f_z$ as a function of magnetron and cyclotron radius. The main contribution to the uncertainty is due to uncertainty in determining $C_4$. Under $\Delta_{trap}$ we also include the effect of uncertainty in the "trap-tilt parameter", $\theta_{mag} = 0.57(5)$ deg., used to obtain $f_m$ from $f_z$ and $f_{ct}$. Under $\Delta_{i-i}$ we list estimates of the shifts to the ratios from perturbation of $f_{ct}$ and $f_z$ of the inner ion due to Coulomb interaction with the outer ion. To lowest order, the outer ion can be treated as a static "ring of charge" that produces additional imperfections to the electrostatic potential that modify $C_4$ and $C_6$, etc. [31]. Shifts to the ratio then result from differences between the two ions in the mode amplitudes of the inner ion (again mainly the axial amplitude following the pi-pulse), and in the parking radius $\rho_{ck}$. Although the estimated shifts are negligible, we assigned experimentally based uncertainties using measurements of the $^{132}$Xe$^{3+}$/$^{129}$Xe$^{3+}$ ratio at different $\rho_{ck}$. Additional shifts to $f_{ct}$ and $f_z$ occur due to the second-order process where the motion of the inner ion non-resonantly excites the outer ion, which then resonantly back-acts on the inner ion [31]. However, even for the close mass-doublet $^{130}$Te$^{3+}$/$^{130}$Xe$^{3+}$ the estimated second-order shift to the ratio is negligible. For $^{130}$Te$^{3+}$/$^{130}$Xe$^{3+}$ we have also included an estimate of the shift (in fact small due to cancellations) due to the drives applied at $f_{ct}$ or $f_{cc}$ of the inner ion exciting the outer ion, hence affecting the ion-ion interaction. Under $\Delta_{fz}$ we give the (significant) systematic corrections we applied to our measured ratios to allow for shifts to the axial frequency due to the "frequency-pushing" interaction of the ion with the resonant detection circuit [29,30], and due to small temporal drifts in the trap voltage, which are not the same for the two ions [36]. Finally, we note that systematic shifts to the ratio due to the ions' image charges in the trap electrodes, and also due to a possible m/q-dependence of the ions' equilibrium positions, are both negligible here. Overall, the largest contribution to the uncertainty in the ratios was the statistical error from the simultaneous fits to the $f_c$ measurements, which was mainly due to magnetic field variation.

*Atomic masses of $^{130}$Te, $^{130}$Xe and $Q_{\beta\beta}$ ($^{130}$Te):* We first convert the cyclotron frequency ratios into mass differences between neutral atoms. To do this we account for the mass of the missing electrons, and the ionization and chemical binding energies which were obtained from Refs. [37-39]. The mass differences corresponding to the ratios in Table I are given in Table II. We note our value for the $^{132}$Xe – $^{129}$Xe mass difference is in excellent agreement with the value 2.999 374 228(6) u (statistical error only), obtained from measurements of $M(^{129}$Xe) and $M(^{132}$Xe) using single ions in Ref. [30].

TABLE II. Mass difference equations corresponding to the ratios given in Table I. The statistical, systematic and total errors are shown in parentheses.

| Ion pair | Mass Difference | Result (u) |
|---|---|---|
| $^{130}$Xe$^{3+}$/$^{129}$Xe$^{3+}$ | $^{130}$Xe – $^{129}$Xe | 0.998 728 483(10)(5)(12) |
| $^{130}$Te$^{3+}$/$^{129}$Xe$^{3+}$ | $^{130}$Te – $^{129}$Xe | 1.001 441 885(10)(5)(12) |
| $^{132}$Xe$^{3+}$/$^{130}$Xe$^{3+}$ | $^{132}$Xe – $^{130}$Xe | 2.000 645 724(11)(7)(14) |
| $^{132}\text{Xe}^{3+}/^{129}\text{Xe}^{3+}$ | $^{132}\text{Xe} - ^{129}\text{Xe}$ | 2.999 374 229(9)(9)(13) |
| $^{130}\text{Te}^{3+}/^{130}\text{Xe}^{3+}$ (1 ion) | $^{130}\text{Te} - ^{130}\text{Xe}$ | 0.002 713 416(33)(4)(34) |
| $^{130}\text{Te}^{3+}/^{130}\text{Xe}^{3+}$ (2 ion) | $^{130}\text{Te} - ^{130}\text{Xe}$ | 0.002 713 402(24)(9)(26) |
| $[^{130}\text{Te}^{3+}/^{130}\text{Xe}^{3+}]\,(^{129}\text{Xe}^{3+})$ | $^{130}\text{Te} - ^{130}\text{Xe}$ | 0.002 713 402(13)(3)(14) |

The data in Table II is intended for use in global, least-squares mass evaluations. Here, for simplicity, we obtain the absolute masses of $^{130}$Te and $^{130}$Xe using only the mass differences in the first three rows, together with the masses of $^{129,132}$Xe in ref. [30], which we treat as reference masses. For $^{130}$Xe we obtain two values from the ratios $^{132}\text{Xe}^{3+}/^{130}\text{Xe}^{3+}$ and $^{130}\text{Xe}^{3+}/^{129}\text{Xe}^{3+}$. We take the weighted average, linearly propagating the systematic uncertainty and the uncertainties in the reference masses. These are compared with values from the AME in Table III.

TABLE III. Atomic masses of $^{130}$Xe and $^{130}$Te obtained from the different ratios, and their weighted averages, compared with previous values.

| Atom | Source | Atomic Mass (u) |
|---|---|---|
| $^{130}$Xe | $^{130}\text{Xe}^{3+}/^{129}\text{Xe}^{3+}$ | 129.903 509 342(16) |
| | $^{132}\text{Xe}^{3+}/^{130}\text{Xe}^{3+}$ | 129.903 509 362(17) |
| | Average | 129.903 509 351(15) |
| | AME [24] | 129.903 508 0(8) |
| $^{130}$Te | $^{130}\text{Te}^{3+}/^{129}\text{Xe}^{3+}$ | 129.906 222 744(16) |
| | AME [24] | 129.906 224 4(21) |

Using the mass difference given in the last row of Table II only, *i.e.* from the ratios with respect to $^{129}\text{Xe}^{3+}$, and 1 u = 931.494 043(80) MeV/$c^2$ [40], we determine the $^{130}$Te – $^{130}$Xe double-beta-decay $Q$-value to be 2527.518(13) keV. Because their uncertainties are larger (and uncertain), we let the other two values for this mass difference remain as checks.

*Conclusions:* Using precision, cryogenic Penning-trap techniques with two, simultaneously-trapped, triply-charged ions, we have measured the atomic masses of $^{130}$Te and $^{130}$Xe to better than 0.2 ppb fractional precision. The mass-energy difference $[M(^{130}\text{Te}) - M(^{130}\text{Xe})]c^2$, equal to the $Q$-value for the double-beta-decay of $^{130}$Te, has been obtained with a 1σ uncertainty of 13 eV. The individual mass measurements of $^{130}$Xe and $^{130}$Te improve upon results in the AME2003 [24] by factors of ~50 and ~130 respectively. Our new value for $Q_{\beta\beta}(^{130}\text{Te})$ agrees with that obtained by the Manitoba group [25], but is a factor of 100 more precise. With more than adequate precision it provides the location of the peak in the total-electron-energy spectrum essential to searches for the neutrino-less double-beta decay of $^{130}$Te.

*Acknowledgements:* We are grateful to D. E. Pritchard and associates, particularly S. Rainville and J.K. Thompson, for the original development of the mass spectrometer and for enabling its relocation to Florida State University. Support was provided by the National Science Foundation under NSF-PHY0652849 and NSF-PHY0500337, and by the National Institute of Standards and Technology PMG program.